# An infrared, Raman, and X-ray database of battery interphase components

Lukas Karapin-Springorum[1,2], Asia Sarycheva[1], Andrew Dopilka[1], Hyungyeon Cha[1], Muhammad Ihsan-Ul-Haq[1], Jonathan M. Larson[1,3]*, and Robert Kostecki[1]*

1. Energy Storage & Distributed Resources Division, Lawrence Berkeley National Laboratory, Berkeley, California, 94720, USA
2. Department of Physics and Astronomy, Pomona College, Claremont, California, 91711, USA (Current Address)
3. Department of Chemistry and Biochemistry, Baylor University, Waco, Texas, 76798, USA (Current Address)
*These authors jointly supervised this work
corresponding authors: Robert Kostecki (R_Kostecki@lbl.gov), Jonathan M. Larson (Jonathan_Larson@Baylor.edu)

## Abstract

Further technological advancement of both lithium-ion and emerging battery technologies can be catalyzed by an improved understanding of the chemistry and working mechanisms of the solid electrolyte interphases (SEIs) that form at electrochemically active battery interfaces. However, collecting and interpreting spectroscopy results of SEIs is difficult for several reasons, including the chemically diverse composition of SEIs. To address this challenge, we herein present a vibrational spectroscopy and X-ray diffraction data library of ten suggested SEI chemical constituents relevant to both lithium-ion and emerging battery chemistries. The data library includes attenuated total reflectance Fourier transform infrared spectroscopy, Raman spectroscopy, and X-ray diffraction data, collected in inert atmospheres afforded by custom designed sample holders. The data library presented in this work (and online repository) alleviates challenges with locating related work that is either diffusely spread throughout the literature, or is non-existent, and provides energy storage researchers streamlined access to vital SEI-relevant data that can catalyse future battery research efforts.

## Background & Summary

Energy storage technologies play a large and growing role in humans' lives and will continue to do so for the foreseeable future. Currently, for many applications, rechargeable lithium-ion batteries (LIBs) have requisite energy and power densities as well as adequate cycle and calendar life at reasonable cost relative to other energy storage technologies.[1,2] As a result, LIBs (with various cathode materials) are used extensively in consumer electronics,[3] deployed in electric vehicles,[4] and are beginning to be used in the electric grid to store energy from intermittent renewable sources.[1,5,6] That said, there is still need for a number of innovations to increase safety and recyclability, reduce cost, and optimize the performance of LIBs to enable additional applications (e.g. aviation, long duration storage) while using materials that minimize supply chain and ethical mining concerns.

From a basic science perspective, many key performance characteristics of LIBs are enabled by the natural formation of solid electrolyte interphases (SEIs) at electrochemically active interfaces. This is because SEIs regulate mass and charge transfer across the interface, as well as the degree of passivation at the interface.[7] A critically important SEI in modern LIBs is the SEI that forms at the interface between the electrolyte and graphite anode, which conducts lithium ions – enabling ion intercalation and deintercalation during charging and discharging – but is an electrical insulator, mitigating deleterious side reactions.[8-10] Fortuitously, the anodic SEI grown during initial formation cycling is quite stable.[11]

It is widely believed that the performance and lifetime of state-of-the-art LIBs and next-generation variants (e.g. those that utilize novel electrochemical intercalation chemistries or the conversion or plating of lithium) can be substantially improved through targeted engineering efforts informed by an improved understanding of the basic structure, chemistry, and working mechanisms of the SEI.[12-15] However, it is difficult to develop an understanding of fundamental structure-function relationships in SEIs in part because they are reactive, chemically heterogeneous, difficult to isolate and sense (being extremely thin, on the order of tens of nanometers thick), and because they are buried between dissimilar materials.[13] Further, the composition of SEIs can depend on manufacturing procedures, cycling conditions, and other physical circumstances which are not always standardized.[8,10,13]

Fourier transform infrared spectroscopy (FTIR) and Raman spectroscopy are commonly used to characterize SEI chemistry and structure, while X-ray diffraction (XRD) can characterize SEI crystallinity, which influences ionic conductivity.[10,13] However, it is often difficult to interpret the data generated by these techniques because of a large mix of contributions to the data, including those from numerous chemical components of the SEI, as well as from other parts of the electrochemical cell (like active electrode materials and current collectors). By various estimates, the SEI has more than ten unique chemical constituents.[13,16] As a result, reference measurements of individual candidate compounds are often used to identify the multiple chemicals contributing to complex spectra. Although some spectral databases exist to assist in this,[17,18] access to these can be prohibitively expensive, or lack needed data. Furthermore, researchers can, of course, collect reference spectra themselves or look to the literature. However, the former is time consuming and possibly costly, and the latter can be challenging for the following reasons. First, published measurements of oxygen- and water-reactive compounds may include varying degrees of contribution from unwanted reaction products that may obscure features of interest. Second, the desired data is commonly difficult to find, being included in articles unrelated to SEI characterization, published in supplemental information sections, or plotted along other data. Finally, if appropriate data is found in the literature, it is not usually offered in a digitized form for streamlined use.

In order to reduce these practical challenges faced by energy storage researchers in the identification of SEI constituents, we present this work, with corresponding online data library posted on Dryad, of measurements of pristine and unreacted SEI components relevant to current and emerging LIB technologies, using attenuated total reflectance FTIR (ATR-FTIR), Raman, and XRD characterization instruments. These compounds are lithium acetate, lithium carbonate, $^6$lithium fluoride, $^7$lithium fluoride, lithium hydride, lithium hexafluorophosphate, lithium oxide, manganese(II) fluoride, nickel(II) fluoride, and polyethylene oxide (PEO). The online repository contains both the final and raw data, as well as the backgrounds removed. Our results both confirm and expand upon those already in the literature. This work, and connected data library, will aid researchers in their efforts to more efficiently identify, or exclude, potential SEI components via data analyses of complex spectra, and thus facilitate basic research of electrochemical interfaces of vital importance to battery materials science.

## Methods

Each of the three empirical characterization methods used – FTIR, Raman, and XRD – are detailed in the forthcoming subsections. Additionally, the protocols employed to ensure chemical compounds remained in an inert environment during storage, transfer, and characterization will also be briefly summarized, and fully validated in the Technical Validation section. Prior to any characterization, all pristine compounds were stored in an argon glovebox with base oxygen and water concentrations of ~0.1 ppm and ~0.5 ppm, respectively, and the sources and purities of the studied chemicals are provided in Table 1.

**Table 1**
Source and purity of compounds used for measurements presented in this paper

| Chemical Name | Chemical Formula | Purity | Supplier | CAS Number |
|---|---|---|---|---|
| Lithium acetate | $CH_3COOLi$ | 99.9% | Sigma Aldrich | 546-89-4 |
| Lithium carbonate | $Li_2CO_3$ | 99.999% | Sigma Aldrich | 554-13-2 |
| [6]Lithium fluoride | [6]LiF | 99%[a] | Sigma Aldrich | 14885-65-5 |
| [7]Lithium fluoride | [7]LiF | 99.99%[b] | Sigma Aldrich | 7789-24-4 |
| Lithium hydride | LiH | 95% | Sigma Aldrich | 7580-67-8 |
| Lithium hexafluorophosphate | $LiPF_6$ | 98% | Thermo Fisher Scientific | 21324-40-3 |
| Lithium oxide | $Li_2O$ | 99.5% | Thermo Fisher Scientific[c] | 12057-24-8 |
| Manganese(II) fluoride | $MnF_2$ | 99% | Thermo Fisher Scientific | 7782-64-1 |
| Nickel(II) fluoride | $NiF_2$ | 97% | Thermo Fisher Scientific | 10028-18-9 |
| Polyethylene oxide | $H(OCH_2CH_2)_nOH$ | 100% | Thermo Fisher Scientific | 25322-68-3 |

a) Advertised 95 atom percent [6]Li, with total compound purity of 99%
b) Advertised average molecular weight of 25.94 implies 92.6 atom percent [7]Li, with total compound purity of 99.99%
c) FTIR data taken using powder from American Elements (99.5%, LI-OX-025M-P)

**ATR-FTIR Spectroscopy.** ATR-FTIR spectra were collected using a Shimadzu IRTracer-100 instrument with an IRIS single reflection diamond accessory from 370 to 4000 $cm^{-1}$ at a spectral resolution of 2 $cm^{-1}$. The instrument was housed in a nitrogen-filled glovebox with an oxygen concentration below 20 ppm. After being transferred into the ATR-FTIR enclosure in sealed vials, compounds were immediately placed on a clean diamond crystal for the ATR-FTIR measurement. This transfer approach was effective at minimizing unwanted reactions, as described in detail in the Technical Validation section below. Most compounds were measured using an average of 512 individual spectra ($CH_3COOLi$, $Li_2CO_3$, [7]LiF, [6]LiF, $Li_2O$, PEO) to maximize the signal to noise ratio, while only 50 spectra were accumulated for some of the more reactive compounds (LiH, $LiPF_6$, $MnF_2$, $NiF_2$) to minimize acquisition time and thereby reduce the likelihood of undesired reactions.

**Raman Spectroscopy.** Raman spectra were collected using a 2 cm-square and 5 mm thick custom-made polyether ketone (PEEK) cell with an optical window (2.5 cm-square and 1 mm thick glass microscope slide). An illustration of the cell is provided in Fig. 1a, and it kept samples in an inert argon environment during measurement. Prior to cell assembly, PEEK cell bodies and glass slides were sequentially sonicated with acetone and isopropyl alcohol, and baked at 40°C for at least 4 hours, before being transferred into the glovebox for assembly. After a cell was assembled, it was enclosed within a heat-sealed bag, before being transferred to a Renishaw Qontor Raman microscope where Raman spectroscopy was conducted. A 488 nm excitation laser used, along with a spectral range of 100 to 3200 $cm^{-1}$, 25 acquisitions, and laser power ranging from 1 to 10 mW. Unwanted contributions to the Raman spectra from the glass optical window were avoided by focusing the laser on the surface of the compounds.

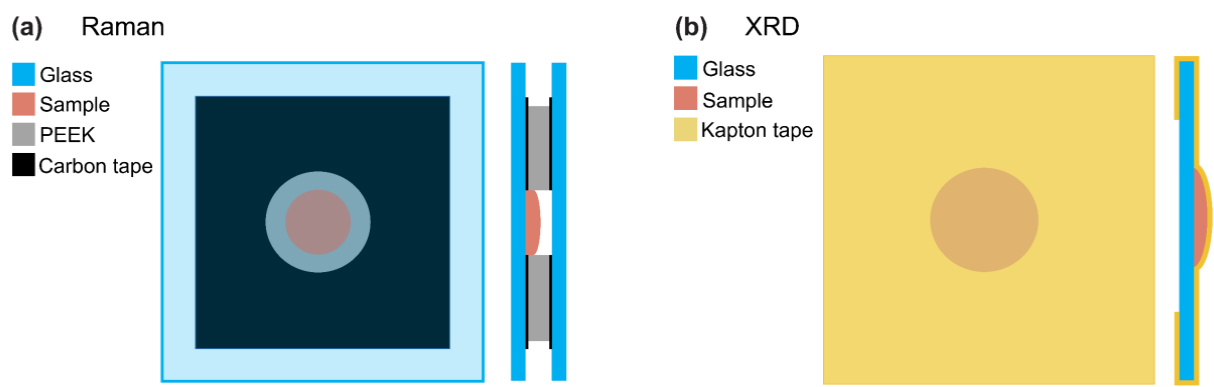

Fig. 1. Cell designs for (a) Raman and (b) XRD measurements.

**X-Ray Diffraction.** Samples for XRD measurements were similarly assembled in the argon glovebox. Each compound was placed on a clean 2.5 cm-square and 1 mm thick glass microscope slide (cleaned and dried using the method described above) and covered with several sealing overlayers of polyimide tape (Kapton, Ted Pella, silicone adhesive, 70 μm thick) before being heat-sealed in individual plastic bags. These sealed samples were then transferred to a Bruker Phaser D2 instrument to collect X-ray diffraction patterns over a 2θ range of 10 to 90 degrees, acquisition time of 0.2 seconds per step, and a step size of 0.02 degrees per step. All samples remained in their sealed bags until right before the start of measurement. XRD patterns were collected through the tape, rather than through glass, to prevent significant XRD contributions from the glass crystal. The amorphous background from the tape was removed via processing as described below in the Data Processing subsection. Comparison of our measurement of $Li_2O$ to the results of Weber *et al.*[19] provides strong evidence that this approach successfully minimized unwanted reactions (see Technical Validation section).

**Data Processing.** The raw collected data was processed to remove unwanted instrumental and/or background contributions. Polynomials or Gaussians were fitted through the low wavenumber background in ATR-FTIR measurements. Spikes in the Raman spectra attributable to cosmic ray excitation were manually removed from Raman spectra. Subsequently, Raman spectra were then processed through the subtraction of Gaussian and/or polynomial fits to eliminate background contributions arising from a number of phenomena (e.g. fluorescence, glass effects, surface roughness). Our instrument generated raw Raman data with unevenly spaced wavenumber values, so an interpolation was performed to transform the data. Gaussian fits to determine peak positions from the data before and after interpolation confirmed transformation did not affect spectral feature shapes and canters. Gaussian fits were used to subtract out an amorphous background in XRD measurements generated by the Kapton tape. This background between 10 and 30 degrees appeared in all measurements through Kapton tape (but not in control measurements of bare metal foils) and was lower than the first diffraction peak of most compounds, allowing for a consistent background subtraction for measurements of the few compounds (like lithium acetate) where the first diffraction peak was found below 30 degrees. For all data types, Fourier filtering was applied to reduce high frequency noise (with care taken to avoid distorting spectral features) and small vertical offsets were used in some cases to align the final baseline near zero. All data was normalized to take on values from 0 to 1.

**Mode Assignment and Notation.** One of the goals and intended contributions of this work is to synthesize existing knowledge about the vibrational modes of the studied compounds, and we make these identifications on plotted data in figures and in accompanying tables. We standardize the notation used to identify peaks where possible; we provide the type of vibrational mode, the bond or functional group that generates the peak, and the symmetry of the vibration. Where this is not possible, we adopt the notation that has been used in the literature that we direct the reader to. References are made to resources that provide additional details regarding the assignment of vibrational modes to spectral features. Unless otherwise noted, the Greek letters are used to describe vibrational modes according to the following associations: $\omega$, wagging; $\delta$, bending; $\nu$, stretching; $\rho$, rocking; $\tau$, twisting. Subscripts "a" and "s" refer to asymmetric and symmetric modes, respectively. Similarly, the subscripts "i" and "o" refer to in-plane and out-of-plane modes, respectively. In the following subsections, we present the data organized according to alphabetical order of the compounds in question.

**Lithium Acetate - $CH_3COOLi$.** Lithium acetate has been identified as an SEI component in Li-$O_2$ batteries[20] and batteries with silicon anodes.[21] The FTIR spectrum in Fig. 2a is in agreement with spectra reported elsewhere.[22-26] We provide assignments for most peaks in the Fig., and in Table 2. The source of the peaks at 1421, 1568, and 1610 $cm^{-1}$ remains unknown, although it has been proposed that they are generated by vibrational modes of the carboxylate group ($COO^-$).[22] Our data shows the presence of peaks at 503, 621, and 657 $cm^{-1}$ which have occasionally been reported in the literature.[24,25] There is some disagreement in the literature as to the shape of the peak near 1600 $cm^{-1}$, and our results agree best with those of Ross[26] and Beyer *et al*.[23] The Raman spectrum in Fig. 2b is in good agreement with the literature, where more detailed peak identifications (as well as empirical peak locations for lithium acetate dihydrate and the free acetate ion) can be found.[27,28] Peak identifications are made in the figure and in Table 2. The XRD pattern in Fig. 2c is in good agreement with that of the anhydrous polymorph obtained from the dehydration of $CH_3COOLi \cdot 2H_2O$.[29] Note that there is substantial variation within the literature as to the XRD pattern of lithium acetate, in part due to the dependency of the crystal structure on the specific hydrate (lithium acetate monohydrate and dihydrate as well as others) used as a precursor[29] and also possibly due to differences in sample preparation procedures that resulted in varying degrees of isolation from water and oxygen.

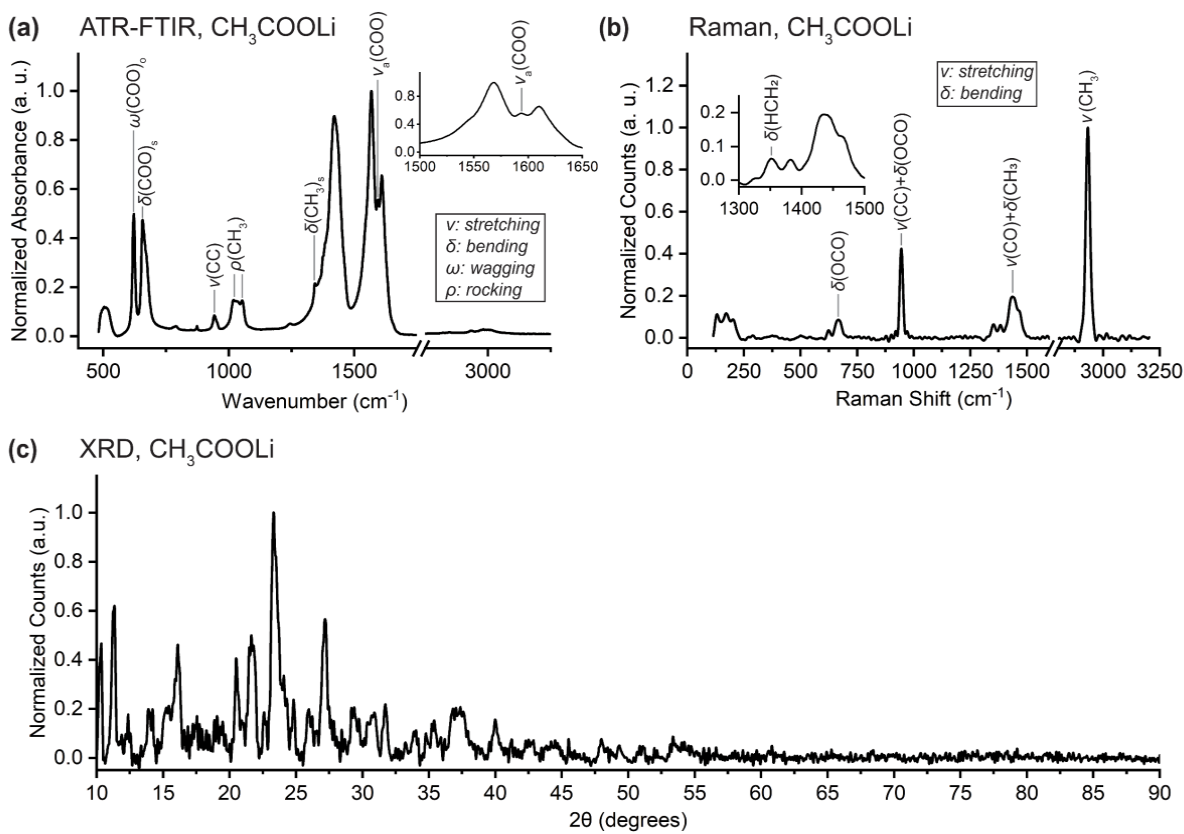

Fig. 2. (a) ATR-FTIR, (b) Raman, and (c) XRD data for lithium acetate ($CH_3COOLi$). Labels for panels (a) are taken from Cadene[30] and for (b) from Sánchez-Carrera and Kozinsky[27] and Ananthanarayanan.[28]

**Table 2**
Peak assignments for FTIR and Raman spectra of lithium acetate

| FTIR | | |
|---|---|---|
| This work (cm$^{-1}$) | Literature[30] (cm$^{-1}$) | Assignment |
| 621 | 621 | $\omega(COO)_o$ |
| 657 | 660 | $\delta(COO)_s$ |
| 942 | 943 | $\nu(CC)$ |
| 1020 | 1031 | $\rho(CH_3)$ |
| 1053 | 1055 | $\rho(CH_3)$ |
| 1343 | 1348 | $\delta(CH_3)_s$ |
| 1594 | 1595 | $\nu_a(COO)$ |

| Raman | | |
|---|---|---|
| This work (cm$^{-1}$) Excitation wavelength: 488 nm | Literature[27] (cm$^{-1}$) Calculated | Assignment |
| 667 | - | $^a\delta(OCO)$ |
| 946 | [942] | $\nu(CC) + \delta(OCO)$ |
| 1352 | [1387] | $\delta(HCH_2)$ |
| 1435 | [1440] | $\nu(CO) + \delta(CH_3)$ |
| 2933 | [2953] | $\nu(CH_3)$ |

Note: Empirical locations of Raman peaks in lithium acetate dihydrate and the free acetate ion available in Ananthanarayanan.[28]
a) Assignment inferred from Ananthanarayanan[28]

**Lithium Carbonate, $Li_2CO_3$.** Lithium carbonate forms in the anodic SEI of LIBs through the decomposition of ethylene carbonate[31] and is also often found on the surface of lithium metal following native passivation.[32] The ATR-FTIR spectrum, reported in Fig. 3a, closely matches most of those previously reported.[23,33-36] Although some spectra are truncated well above 500 cm$^{-1}$, Özer et al.[35] and Pasierb et al.[36] report a strong peak near 500 cm$^{-1}$ and a weaker peak at slightly lower wavenumbers (both assigned to quasi-lattice vibrations[36]), similar to the peaks that we report at 477 and 408 cm$^{-1}$. The Raman spectrum in Fig. 3b is in good agreement with the literature.[36,37] The peaks that we report in the low-wavenumber region (at 128, 156, 194, and 274 cm$^{-1}$) can be attributed to translational or rotational lattice vibrations of the carbonate ion ($CO_3^{2-}$).[38,39] Peak identifications for FTIR and Raman-active vibrational modes are made in the figure and in Table 3. Additional information about vibrational modes can be found in Brooker and Bates,[40] Brooker and Wang,[41] and Hase and Yoshida.[38] The XRD pattern in Fig. 3c confirms peak positions found in the literature and is consistent with a monoclinic crystal structure.[42-44] We report relative peak intensities that differ somewhat from previously reported values, although these values in the literature are generally not consistent with each other. Of note are the smaller peaks between 55 and 60 degrees, whose positions and relative intensities have not been agreed upon and which have been assigned to different crystal planes.[42,44]

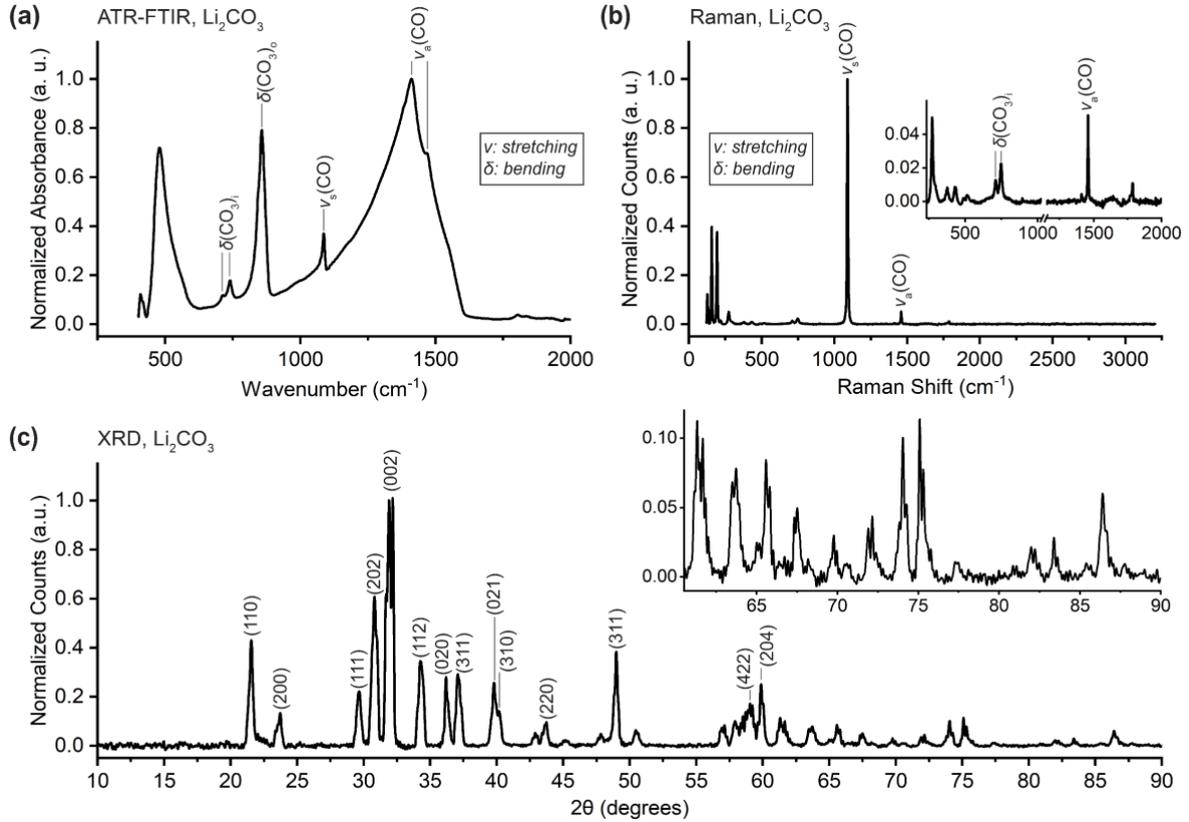

Fig. 3. (a) ATR-FTIR, (b) Raman, and (c) XRD data for lithium carbonate (Li$_2$CO$_3$). Labels are taken from the literature: (a) and (b) Pasierb *et al.*[36] and (c) Chen *et al.*[44]

**Table 3**
Peak assignments for FTIR and Raman spectra of lithium carbonate

| FTIR | | |
|---|---|---|
| This work (cm$^{-1}$) | Literature[36] (cm$^{-1}$) | Assignment |
| 713 | 712 | $\delta(CO_3)_i$ |
| 740 | 748 | $\delta(CO_3)_i$ |
| 857 | 863 | $\delta(CO_3)_o$ |
| 1087 | 1088 | $\nu_s(CO)$ |
| 1412 | 1437 | $\nu_a(CO)$ |
| 1468 | 1503 | $\nu_a(CO)$ |

| Raman | | |
|---|---|---|
| This work (cm$^{-1}$) | Literature[36] (cm$^{-1}$) | Assignment |
| Excitation wavelength: 488 nm | Excitation wavelength: 1064 nm | |
| 710 | 712 | $\delta(CO_3)_i$ |
| 748 | 748 | $\delta(CO_3)_i$ |
| 1090 | 1088 | $\nu_s(CO)$ |
| 1459 | 1458 | $\nu_a(CO)$ |

**Lithium Fluoride - $^7$LiF & $^6$LiF.** Lithium fluoride is a primary component of the initial SEI in LIBs and is produced when LiPF$_6$ is reduced during a reaction with ethylene carbonate.[31,45] Both $^7$LiF and $^6$LiF are naturally occurring to some extent and their vibrational spectra are expected to be influenced by their differing atomic weights.[46,47] The ATR-FTIR spectra in Fig. 4a are similar to each other in that they have broad double peaks in the low wavenumber region. However, the peaks of the lighter $^6$LiF, at 527 and 595 cm$^{-1}$, are shifted approximately 20 cm$^{-1}$ higher than those of $^7$LiF (at 508 and 568 cm$^{-1}$), in accordance with well-established theory about the effect of isotope masses on the frequency of vibrational modes.[46] These features contrast with identifications, including a sharp peak near 800 cm$^{-1}$ attributed to LiF elsewhere, and those in spectra of LiF collected at high temperature while within an argon matrix.[48,49] While Raman data was collected, we do not plot it here because high fluorescence overwhelmed features of interest. The raw data is made available in the Dryad repository. As seen in Fig. 4c, both compounds have XRD patterns that are consistent with their cubic structure and are in accordance with the literature.[50-52] Note that data is often reported only through 80 degrees, excluding the peak at 83 degrees assigned to the 222 crystal plane.[53] We find that the diffraction peaks of $^6$LiF are shifted slightly higher than those of $^7$LiF, which would be qualitatively consistent with expected isotope effects, albeit this shift is somewhat larger then what may be expected.[54] Diffraction peaks for $^7$LiF around 79 and 83 degrees are assigned to the 311 and 222 crystal planes in the literature, and we extend these assignments to $^6$LiF on the basis of their shared cubic crystal structure and very similar unit cell size.

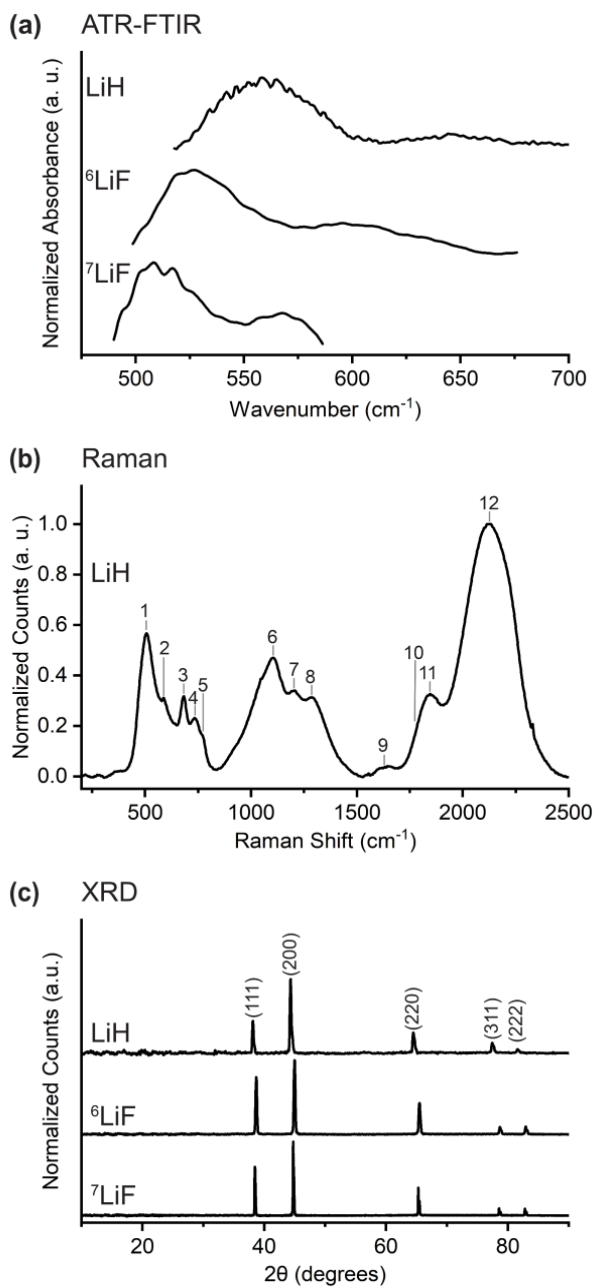

Fig. 4. (a) ATR-FTIR, (b) Raman, and (c) XRD data for lithium fluoride ($^7$LiF and $^6$LiF) and lithium hydride (LiH). Numeric labels in panel (b) correspond to linear combinations of optical and acoustic modes in the LiH crystal as detailed in Tyutyunnik and Tyutyunnik.[55] Labels in panel (c) apply to all three compounds and are from the following sources: LiH labels from Weber *et al*,[56] $^6$LiF labels from Carturan *et al.*,[57] and $^7$LiF labels from Zhang *et al.*[51] and Paterson.[53]

**Lithium Hydride – LiH.** Lithium hydride is a common component of the SEI in lithium-anode batteries and has been identified in lithium dendrites in LIBs, including dendrites composed primarily of LiH that are associated with substantial capacity fading.[58] Importantly, LiH is sometimes misidentified as LiF due to similarities in their chemical properties, underscoring the importance of having quality reference data for both compounds provided in a single location.[59] The ATR-FTIR spectrum in Fig. 4a contains peaks at 558 and 646 cm$^{-1}$; the shift to higher wavenumber relative to the heavier lithium fluoride is consistent with theoretical expectations.[46] The main peak location aligns with a previously reported spectra, although we may be the first to report the weak secondary peak at 648 cm$^{-1}$.[60] The Raman spectra in Fig. 4b reproduces existing spectra and confirms that the sample has not reacted with water to form $Li_2O$ or LiOH.[55,61] Peak assignments are adopted from Tyutyunnik and Tyutyunnik[55] and the peak label "10" indicates the location where they find a modest peak that we do not observe in our data. Raman peak locations and assignments are also provided in Table 4. As seen in Fig. 4c, lithium hydride produces an XRD pattern similar to that of lithium fluoride, likely due to their cubic crystal structures and comparable unit cell sizes. This is consistent with the literature.[56,60,62] The absence of peaks at 33 and 56 degrees (attributed to $Li_2O$) confirms the pristine nature of LiH.[56]

**Table 4**
Peak assignments for Raman spectrum of lithium hydride

| This work (cm$^{-1}$) Excitation wavelength: 488 nm | Literature[55] (cm$^{-1}$) Excitation wavelength: 514.5 nm | Assignment |
|---|---|---|
| 514 | 511 | 1 |
| 592 | 586 | 2 |
| 688 | 687 | 3 |
| 748 | 736 | 4 |
| 779 | 776 | 5 |
| 1094 | 1109 | 6 |
| 1221 | 1211 | 7 |
| 1305 | 1293 | 8 |
| 1649 | 1638 | 9 |
| - | 1736 | 10 |
| 1833 | 1835 | 11 |
| 2142 | 2113 | 12 |
| Note: Detailed peak identifications provided in the reference[55] | | |

**Lithium Hexafluorophosphate - $LiPF_6$.** Lithium hexafluorophosphate is commonly used as a lithium source in liquid electrolytes of LIBs. As a result, residual salt is commonly found on and in the SEI, especially after the evaporation of the volatile components of the electrolyte.[63] The ATR-FTIR spectrum is provided in Fig. 5a. Peak assignments in Fig. 5 and Table 5 adopt the notation that is consistently used in the literature. Note that Pekarek et al.[64] assign the peaks that they observe at 559 and 871 cm$^{-1}$ to $δ_s$(FPF) and $v_s$(PF), respectively. The Raman spectrum in Fig. 5b is in excellent agreement with previously reported values and peak identifications are provided in Table 5.[65-67] The $v_1$ vibrational mode at 771 cm$^{-1}$ has been identified as arising from a P-F symmetric stretching mode.[68] It is worth noting that the normal modes of vibration of the $PF_6^-$ ion are dependent on the coordination of the ion and so can be expected to vary according to the local chemistry of a given SEI.[68] Moreover, partial oxidation of the $PF_6^-$ ion (which can occur upon electrochemical cycling) can also lead to changes in the position of the primary peaks.[65] Presented in Fig. 5c, the peak locations in the XRD pattern are in good agreement with the literature.[65,67,69] Further, we find relative peak intensities that fall in the wide range of results reported by others, with exception of a much stronger peak at 52 degrees than is reported elsewhere. There is some disagreement in the literature as to the crystal structure of $LiPF_6$; Liu et al.[69] suggest a hexagonal crystal structure and Kock et al.[65] use Raman data to propose a cubic crystal structure like that of $NaPF_6$ and $KPF_6$.

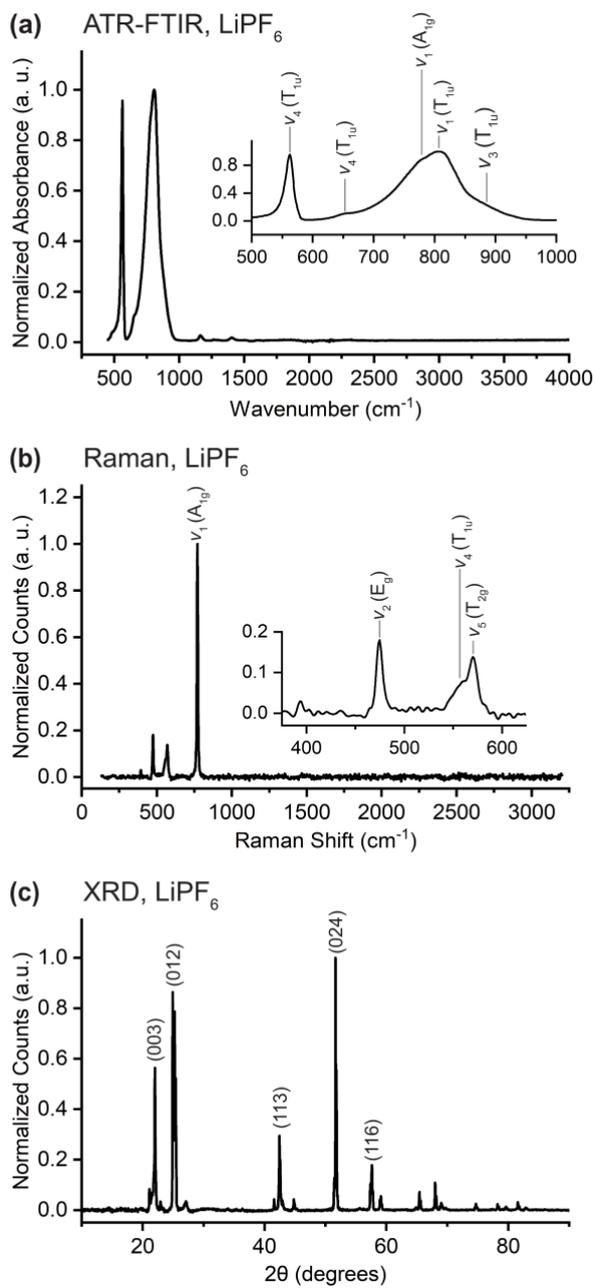

Fig. 5. (a) ATR-FTIR, (b) Raman, and (c) XRD data for lithium hexafluorophosphate (LiPF$_6$). Labels from the literature: (a) and (b) from Kock *et al.*[65] and (c) from Masoud *et al.*[70]

**Table 5**
Peak assignments for FTIR and Raman spectra of lithium hexafluorophosphate

| FTIR | | |
|---|---|---|
| This work (cm$^{-1}$) | Literature[65] (cm$^{-1}$) | Assignment |
| 561 | 560 | $v_4$ (T$_{1u}$) |
| 650 | 646 | $v_4$ (T$_{1u}$) |
| 778 | 775 | $v_1$ (A$_{1g}$) |
| 805 | 798 | $v_1$ (T$_{1u}$) |
| 887 | 869 | $v_3$ (T$_{1u}$) |

| Raman | | |
|---|---|---|
| This work (cm$^{-1}$) | Literature[65] (cm$^{-1}$) | Assignment |
| Excitation wavelength: 488 nm | Excitation wavelength: 1064 nm | |
| 475 | 475 | $v_2$ (E$_g$) |
| 561 | 560 | $v_4$ (T$_{1u}$) |
| 571 | 571 | $v_5$ (T$_{2g}$) |
| 771 | 771 | $v_1$ (A$_{1g}$) |

Note: Notation adopted from Kock *et al.*[65]

**Lithium Oxide - Li$_2$O.** Lithium oxide is believed to form in the SEI as a decomposition product of lithium ethylene dicarbonate[31] and also forms on the surface of lithium metal via native passivation.[32] It can be difficult to obtain pristine spectra of lithium oxide; not only because it reacts readily with water, but also because it is typically offered by vendors with some existing contamination/inpurity.[19] While absorption features were observed at higher wavenumber values, ATR-FTIR data of Li$_2$O is presented here in Fig. 6a below 800 cm$^{-1}$ to focus on Li$_2$O-specific features, which are fairly consistent with sparce reports in the literature (e.g. in the supplemental section of Tian *et al.*[71]). In order to verify infrared spectral features originating strictly from Li$_2$O vibrations, we collected a number of spectra of powders exposed for various amounts of time to gaseous water, CO$_2$, and/or heating at 400 °C (to dehydrate the sample). As a result, features attributable to Li$_2$CO$_3$ (870, 1432, and 1499 cm$^{-1}$)[71-73] and LiOH (3523, 3676 cm$^{-1}$)[74] were identified. These features scaled with exposure to moisture and CO$_2$ at room temperature. However, and interestingly, one absorption feature associated with LiOH (3523 cm$^{-1}$) was found to decrease in intensity with increasing dehydration by way of heating. Thus, and in accord with past reports, the weak peak at ca. 3523 cm$^{-1}$ is assigned to LiOH·H$_2$O,[75,76] while the peak at 3676 cm$^{-1}$ is assigned to the O-H stretching mode in LiOH. Ultimately, all these "impurities" were found above 800 cm$^{-1}$, hence our choice of scale in Fig. 6a.

The Raman spectrum of lithium oxide is provided in Fig. 6b. The feature at 523 cm$^{-1}$ of the main Li-O vibration (in the $F_{2g}$ symmetry group[77]) is congruent with the range of peak positions, from 515 to 529 cm$^{-1}$, reported in the literature[19,78] (including the Supplemental Materials section of Gittleson *et al.*[79]) and precisely confirms the experimental result of Sánchez-Carrera and Kozinsky[27] and Osaka and Shindo.[77] The variance in the previously reported peak locations may be due to differences in the excitation laser wavelength used, as seen in Table 6, although we observe the same peak position when using a 488 and 633 nm excitation lasers. We present the collected XRD pattern, which is in good agreement with data reported elsewhere, in Fig. 6b.[80,81] This pattern is consistent with a cubic crystal structure.

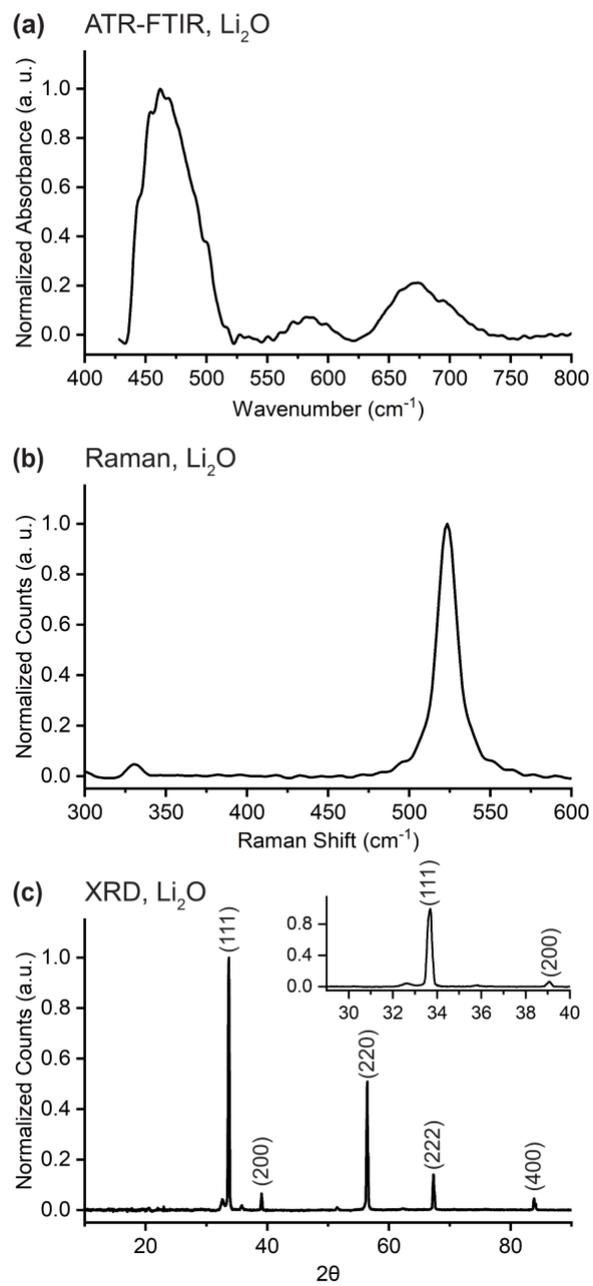

Fig. 6. (a) FTIR, (b) Raman, and (c) XRD data for lithium oxide (Li$_2$O). Labels for panel (c) from the Materials Project.[82]

**Table 6**
Primary Raman peak location of lithium oxide

| Excitation Wavelength (nm) | Primary Peak Location (cm$^{-1}$) | Reference |
| --- | --- | --- |
| 325 | 515 | Sifuentes et al.[78] |
| 488 | 523 | This work |
| 488 | 523 | Osaka and Shindo[77] |
| 514.5 | 523 | Osaka and Shindo[77] |
| 532 | 527 | Weber et al.[19] |
| 633 | 523 | This work |
| 785 | 529 | Gittleson et al.[79] |

**Manganese(II) Fluoride - MnF$_2$.** Manganese(II) fluoride may form in the SEI on manganese-containing metallic glass anodes following reactions with LiPF$_6$ salt in the electrolyte[83] and on other manganese-rich electrodes.[84-86] The ATR-FTIR spectrum presented in Fig. 7a contains peaks similar to those calculated and empirically reported in Scholz and Stösser[87] as seen in Table 7. We find a peak at 575 cm$^{-1}$ that may be generated by the symmetric stretching mode whose position they calculated but which they did not experimentally report.[87] The Raman spectrum in Fig. 7b agrees closely with previously reported values.[88] We provide the symmetry groups of the phonon modes believed to generate the peaks in the low wavenumber region. The peak at 641 cm$^{-1}$ remains unassigned. Seen in Fig. 7c, XRD suggests a tetragonal structure and is in good agreement with published patterns.[89-91] While relative peak intensities are generally similar to those found in the literature, the peak at 50 degrees is somewhat stronger than reported elsewhere.

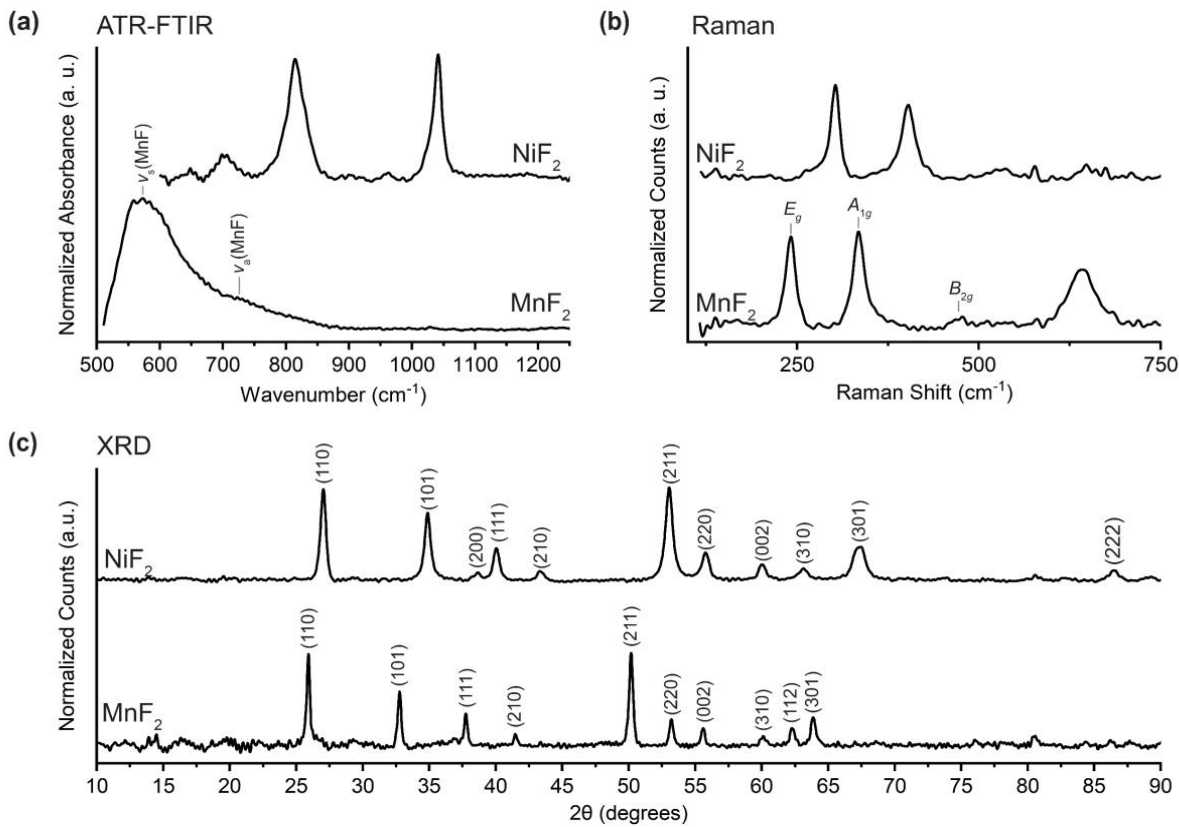

Fig. 7. (a) ATR-FTIR, (b) Raman, and (c) XRD data for manganese(II) fluoride (MnF$_2$) and nickel(II)

fluoride (NiF$_2$). Labels from the literature: (a) Scholz and Stösser,[87] (b) Stavrou et al.,[88] (c) Rui et al.[89] for MnF$_2$, (c) Jiao et al.[92] for NiF$_2$.

**Table 7**
Peak assignments for FTIR and Raman spectra of manganese(II) fluoride

| FTIR | | |
|---|---|---|
| This work (cm$^{-1}$) | Literature[87] (cm$^{-1}$) | Assignment |
| 575 | --- [529-599] | $\nu_s$(MnF) |
| 726 | 700 [673-768] | $\nu_a$(MnF) |

| Raman | | |
|---|---|---|
| This work (cm$^{-1}$) Excitation wavelength: 488 nm | Literature[88] (cm$^{-1}$) Excitation wavelength: 488 nm | Assignment (Phonon Mode)[a] |
| 247 | 245 [233] | E$_g$ |
| 341 | 340 [350] | A$_{1g}$ |
| 476 | 457 [463] | B$_{2g}$ |

Note: Calculated frequencies in square brackets
a) Raman peak identifications refer to phonon modes.

**Nickel(II) Fluoride - NiF$_2$.** Nickel(II) fluoride is believed to form in the SEI on lithium nickel cobalt manganese oxide cathodes in LIBs.[93,94] There is substantial disagreement in the literature as to the correct FTIR spectrum; our data is shown in Fig. 7a and most closely matches that of Tramšek et al.[95] with primary peaks at 817 and 1043 cm$^{-1}$. Note that Tramšek et al. report weak peaks similar to those that we find at 651 and 706 cm$^{-1}$ but that they are more prominent in our data due to our use of a background subtraction. Presented in Fig. 7b, our Raman spectrum shows peaks at 303 and 405 cm$^{-1}$ in good agreement with Ullah et al.[96] The XRD pattern in Fig. 7c closely matches those reported in the literature with respect to both peak locations and relative intensities (with the exception of the strong peak that we report at 53 degrees) and is consistent with the expected tetragonal structure.[92,97,98] Note that the broad peak near 67 degrees has been suggested to be two adjacent peaks stemming from the 301 and 112 crystal planes but we do not have sufficient resolution to make this identification.[97]

**Polyethylene oxide - H(OCH$_2$CH$_2$)$_n$OH.** Polyethylene oxide is a polymer used in polymer electrolytes and as a component in artificial SEIs for anode-free batteries.[99,100] PEO oligomers have also been identified as naturally forming components of the SEI on silicon anodes and are believed to be products of electrolyte reduction.[101] Seen in Fig. 8a, the FTIR spectrum that we present contains the low-wavenumber peaks at 509 and 530 cm$^{-1}$ and the peak at 2876 cm$^{-1}$ reported in some works[102-105] but only faintly shows the broad feature near 3370 cm$^{-1}$ found by some[102-104] but not others.[105,106] Peak locations and vibrational mode assignments are provided in the figure and in Table 8; phase relations of "+" and "-" are provided for coupled coordinates where available but are otherwise replaced with a comma. Additional discussions of vibrational modes can be found in the literature.[107] The Raman spectrum presented in Fig. 8b is in good agreement with the existing literature.[108-110] The XRD pattern in Fig. 8c aligns well with those reported in the literature.[111-113]

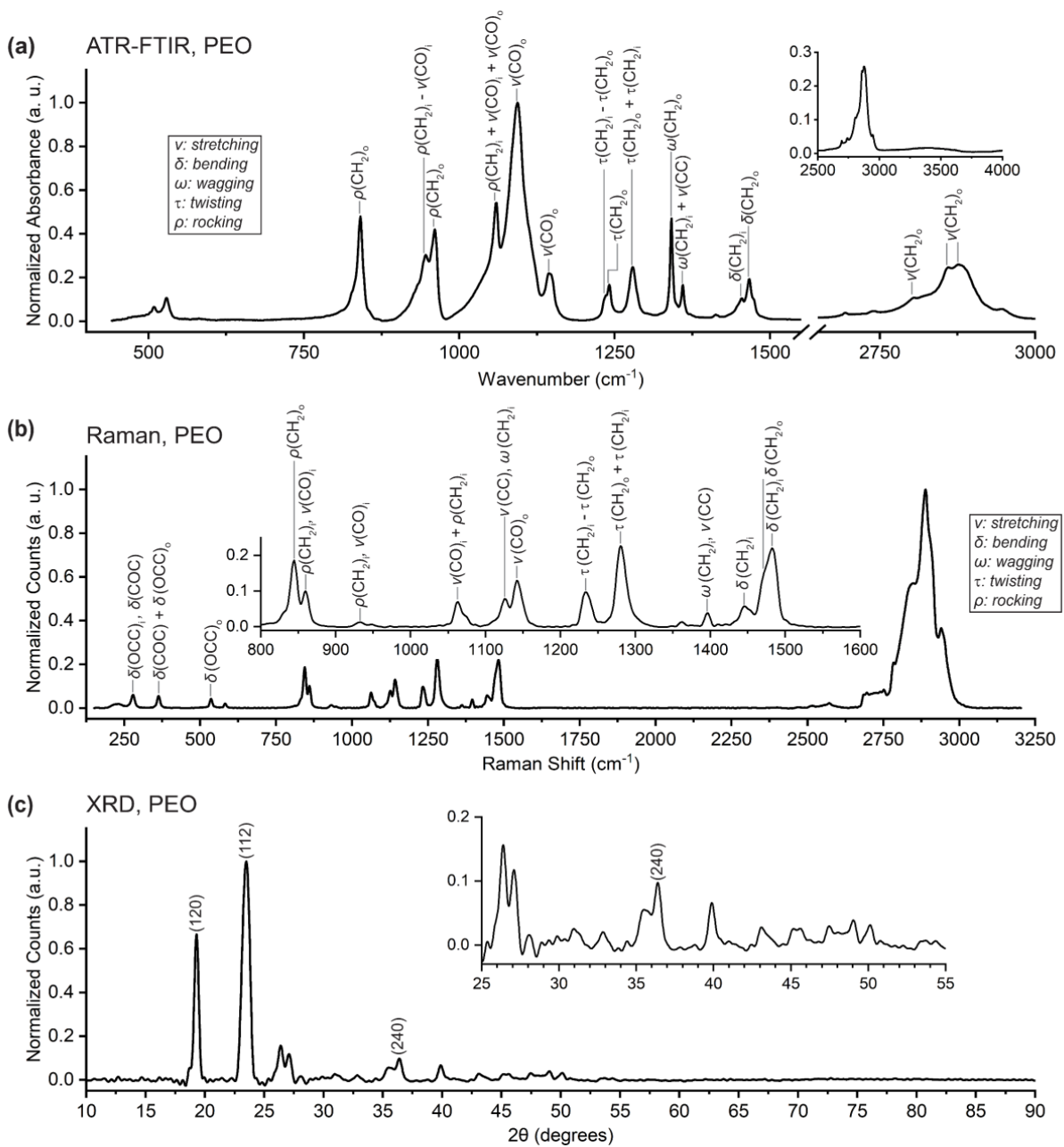

Fig. 8. (a) ATR-FTIR, (b) Raman, and (c) XRD data for polyethylene oxide (H(OCH$_2$CH$_2$)$_n$OH). Labels from the literature: (a) Matsui et al.[114] and Ratna et al.,[115] (b) Matsui et al.[114] and (c) Telfah et al.[116] Phase relations of "+" and "-" are provided for coupled coordinates where available but are otherwise replaced with a comma.

**Table 8**

Peak assignments for FTIR and Raman spectra of polyethylene oxide

| FTIR | | |
|---|---|---|
| This work (cm$^{-1}$) | Literature[114,115] (cm$^{-1}$) | Assignment |
| 841 | 844 [856] | $\rho(CH_2)_o$ |
| 946 | 947 [924] | $\rho(CH_2)_i - \nu(CO)_i$ |
| 961 | 958 [884] | $\rho(CH_2)_o$ |
| 1060 | 1060 [1033] | $\rho(CH_2)_i + \nu(CO)_i + \nu(CO)_o$ |
| 1094 | 1103 [1061] | $\nu(CO)_o$ |
| 1144 | 1147 [1161] | $\nu(CO)_o*$ |
| 1236 | 1234 [1250] | $\tau(CH_2)_i - \tau(CH_2)_o$ |
| 1241 | 1240 [1280] | $\tau(CH_2)_o$ |
| 1279 | 1278 [1282] | $\tau(CH_2)_o + \tau(CH_2)_i$ |
| 1342 | 1342 [1386] | $\omega(CH_2)_o$ |
| 1360 | 1358 [1354] | $\omega(CH_2)_i + \nu(CC)$ |
| 1454 | 1448 [1470] | $\delta(CH_2)_i$ |
| 1467 | 1466 [1474] | $\delta(CH_2)_o$ |
| 2806 | 2750-3000 | $\nu(CH_2)_o$ |
| 2861 | 2750-3000 | $\nu(CH_2)_o$ |
| 2876 | 2750-3000 | $\nu(CH_2)_o$ |

| Raman | | |
|---|---|---|
| This work (cm$^{-1}$) | Literature[114] (cm$^{-1}$) | Assignment |
| Excitation wavelength: 488 nm | Excitation wavelength: 435.8 nm | |
| 279 | 274 [270] | $\delta(OCC)_i, \delta(COC)$ |
| 363 | 359 [366] | $\delta(COC) + \delta(OCC)_i$ |
| 536 | 531 [501] | $\delta(OCC)_o*$ |
| 844 | 844 [856] | $\rho(CH_2)_o$ |
| 860 | 859 [876] | $\rho(CH_2)_i, \nu(CO)_i$ |
| 933 | 930 [924] | $\rho(CH_2)_i, \nu(CO)_i$ |
| 1062 | 1066 [1093] | $\nu(CO)_i + \rho(CH_2)_i$ |
| 1126 | 1130 [1138] | $\nu(CC), \omega(CH_2)_i$ |
| 1142 | 1147 [1161] | $\nu(CO)_o*$ |
| 1231 | 1237 [1250] | $\tau(CH_2)_i - \tau(CH_2)_o$ |
| 1280 | 1283 [1282] | $\tau(CH_2)_o + \tau(CH_2)_i$ |
| 1396 | 1398 [1381] | $\omega(CH_2)_i, \nu(CC)$ |
| 1446 | 1447 [1470] | $\delta(CH_2)_i$ |
| 1470 | 1474 [1474] | $\delta(CH_2)_o$ |
| 1480 | 1483 [1473] | $\delta(CH_2)_i$ |

Note: Calculated frequencies in square brackets. Phase relations of "+" and "-" are provided for coupled coordinates where available but are otherwise replaced with a comma.
*Alternative identification is made in Tadokoro *et al*.[107]

## Data Records

All data presented or discussed herein can be found in the Dryad repository associated with this work. The data files are in a .xlsx format which can be opened using Excel (including the free Excel Viewer) and Google Sheets among other applications. These files can be saved out in a .csv or other format as desired for use in other software (including data processing and plotting programs). One Excel workbook file is provided for each characterization technique (ATR-FTIR, Raman, and XRD). Each sheet contains data for one compound, whose identity is indicated by the chemical formula and written name. Raw and final data are provided, and the ATR-FTIR and XRD data files also contain an additional sheet with all final data on a single x axis. Note that final Raman data for $^6$LiF and $^7$LiF are omitted as discussed in the Methods section, and that raw Raman data has uneven x-axis spacing so had to be interpolated to apply the Fourier filter. There are therefore different wavenumber axes for the raw and final Raman data; the axes are provided in each sheet and are visually offset by an empty column. The .txt file in the repository folder contains a duplicate of the information provided in this section.

## Technical Validation

A primary consideration when designing the procedures outlined in the Methods section was to ensure the purity of the compounds that we measured. Our success in doing so was confirmed through the comparison of our data to literature results of both reacted and pristine compounds of interest. Comparison of the ATR-FTIR spectrum of LiPF$_6$ (shown in Fig. 5a) to the literature demonstrates our measurement procedure's efficacy in preventing unwanted side reactions; the LiPF$_6$ is demonstrably unreacted.[65,67] The efficacy of our Raman measurement protocols is supported by the lithium hydride spectrum in Fig. 4b; the absence of peaks at 523 cm$^{-1}$ (from Li$_2$O) and between 250 and 350 cm$^{-1}$ (attributed to LiOH)[61] provides strong evidence that this highly oxygen- and water-sensitive compound has not formed its common reaction products. Further, the quality of the XRD data is confirmed through analysis of our lithium oxide data (in Fig. 6c); there are strong peaks generated by Li$_2$O with almost negligible contributions from the 101 plane of LiOH at ~32 degrees (which have been suggested to arise from impurities in the as-delivered powder[19]) no detectable contributions from the 110 plane of LiOH·H$_2$O at ~36 degrees, indicating that the sample was well protected from oxygen and had no detectable exposure to water.[19] Further, the absence of peaks at 33 and 56 degrees in the XRD pattern of LiH in Fig. 4c confirms the absence of LiOH contamination in the LiH.[56]

Our data also closely aligns with previously reported spectra and patterns (where these are available) for FTIR (lithium acetate[23,26], lithium carbonate[35,36], lithium hydride[60], lithium hexafluorophosphate[65,67], nickel(II) fluoride[95], polyethylene oxide[106]), Raman (lithium acetate[27,28], lithium carbonate[36,37], lithium hydride[55], lithium hexafluorophosphate[65-67], lithium oxide[27,77], manganese(II) fluoride[88], nickel(II) fluoride[96], polyethylene oxide[108-110]), and XRD (lithium acetate[29], lithium carbonate[42-44], $^6$lithium fluoride[52], $^7$lithium fluoride[50,51], lithium hydride[56,60,62], lithium hexafluorophosphate[65,67,69], lithium oxide[19,80,81], manganese(II) fluoride[89-91], nickel(II) fluoride[92,97,98], polyethylene oxide[111-113]) measurements.

## Acknowledgments
We kindly acknowledge sources that financially supported this work. Funding to support this work was provided by the Assistant Secretary for Energy Efficiency and Renewable Energy, Vehicle Technologies Office, under the Advanced Battery Materials Research (BMR) Program, of the U.S. Department of Energy under Contract No. DE-AC02-05CH11231. Additionally, funding to support this work was also provided by the Energy & Biosciences Institute through the EBI-Shell program.

## Author contributions
J.M.L. and R.K. supervised the work. J.M.L. conceived of the project concept. L. K.-S. prepared the samples and collected a vast majority of the data while being supported by A.S., A.D., M. I.-U.-H., H.C., and J.M.L. with sample storage and preparation advice, miscellaneous data collection, and training. L. K.-S. wrote and prepared the manuscript draft with J.M.L.'s supervision and guidance. All authors contributed to the interpretation, conclusions, and preparation of the final manuscript.

## Competing interests
The authors declare no competing interests.